\documentclass[twoside,leqno,twocolumn]{article}  
\usepackage{ltexpprt} 

\usepackage[latin1]{inputenc}
\usepackage{amstext}
\usepackage{graphicx}
\usepackage{amsfonts}
\usepackage{latexsym}
\usepackage{amssymb}
\usepackage{amsmath}
\usepackage{floatflt}

\newtheorem{mytheorem}{Theorem}[section]
\newtheorem{myalg}[mytheorem]{Algorithm}

\newcommand{\PR}{\mathbf{P}}

\newcommand{\R}{\mathbb{R}}
\newcommand{\Z}{\mathbb{Z}}

\title{Adaptive Dynamics of Realistic Small-World Networks} 

\author{Olof Mogren\thanks{Department of
    Computer Science and Engineering, Chalmers University of Technology
and Göteborg University}
  \and Oskar Sandberg\thanks{Department of Mathematical Sciences,
    Chalmers University of Technology and Göteborg University, 412 96
    Göteborg, Sweden. ossa@math.chalmers.se \emph{Corresponding author.} } \and Vilhelm
  Verendel\thanks{Department of
    Computer Science and Engineering, Chalmers University of
    Technology and Göteborg University} \and Devdatt Dubhashi\thanks{Department of
    Computer Science and Engineering, Chalmers University of
    Technology and Göteborg University}}

\date{\today}

\begin{document}

\maketitle

\begin{abstract}
  Continuing in the steps of Jon Kleinberg's and others celebrated
  work on decentralized search in small-world networks, we conduct an
  experimental analysis of a dynamic algorithm that produces
  small-world networks. We find that the algorithm adapts robustly to
  a wide variety of situations in realistic geographic networks with
  synthetic test data and with real world data, even when vertices
  are uneven and non-homogeneously distributed.

  We investigate the same algorithm in the case where some vertices
  are more popular destinations for searches than others, for example
  obeying power-laws. We find that the algorithm adapts and adjusts
  the networks according to the distributions, leading to improved
  performance. The ability of the dynamic process to adapt and create
  small worlds in such diverse settings suggests a possible mechanism
  by which such networks appear in nature.
\end{abstract}

\section{Introduction}

In 1967 Stanley Milgram set out to measure the ``smallness'' of the
world. He wanted to know if it was really true that any two people
could be connected through a short chain of acquaintances. To conduct
this experiment, he gave volunteers living in Omaha, Nebraska, a
letter addressed to a stockbroker from outside Boston, asking them to
forward it to him, with the stipulation that the letter could only
ever pass between people who were on a first name basis. The results
of his experiment were generally seen as proof that we really do live
in a small world -- for the letters that arrived successfully, the
average number of steps was just six.

In mathematics, the idea of the small world has inspired the study of
graph diameter. Roughly speaking, it has been noted that if the edges
of a graph are chosen randomly, then the diameter tends to be
``small'': of the order of $\log n$ where $n$ is the size of the
graph. However, while such a world may be small, this does not
in itself explain the success of Milgram's experiment. In his seminal
paper from 2000 \cite{kleinberg:smallworld}, Jon Kleinberg took an
algorithmic perspective and asked: how is it that it was possible for
people to know whom they should send the letter to so it would arrive
in few steps? After all, the social network is a criss-crossed maze of
connections, of which the participants have no overview.  Kleinberg
showed that for it to be possible, using only local knowledge, to
efficiently forward the message to its destination the graph must have
a particular form. Specifically, the probability that two people are
acquainted must follow a particular power-law relation with the
distance between them. When this is the case, messages can be routed
in a polylogarithmic number of steps, in all other cases it is
exponentially larger. Graphs where routing is efficient have since
been labeled \emph{navigable}.

\subsection{Motivation}

In Kleinberg's original work \cite{kleinberg:smallworld}
\cite{kleinberg:navigation}, his model for the world was a
two-dimensional grid, where people knew their $k$-nearest neighbors,
and had $r$ random long-range contacts in the network. The
distribution of these \emph{shortcuts} is what determines the
navigability of the graph. Later works have extended the model to more
general and more realistic settings. In his PhD thesis, David
Liben-Nowell \cite{nowell:phd} studied a real-world social network of
people from the United States connected over the Internet. He found
that this network was navigable, and could be made to fit with
Kleinberg's theory, but only after adjustments had been made to take
into account the highly non-homogeneous geographical distribution of
the population. While his work gives hints as to in what situations
the unadjusted model fails, the criteria for this have not been
characterized.  Several works have explored this more general relation
in other contexts \cite{kleinberg:dynamics} \cite{fraigniaud:greedy}
\cite{duchon:anygraph}.

\begin{figure}[t]
  \begin{center}
    \includegraphics[width=8cm]{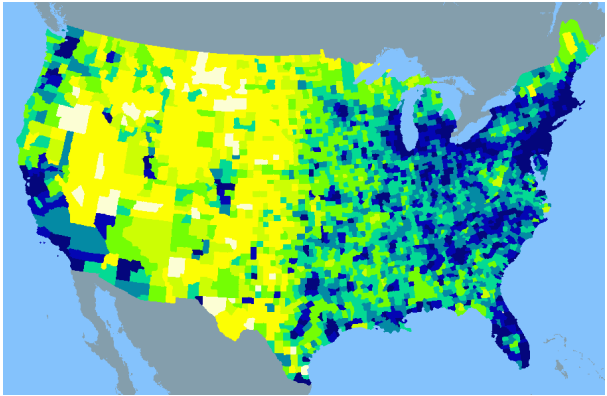} 
  \end{center}
  \caption{In the real world, populations are in-homogeneously
    distributed. Here the population density of the United States of
    America by county.}
  \label{fig:usapop}    
\end{figure}

Another question that is raised by any attempt to apply Kleinberg's
ideas to the real world, is understanding why social networks should
be navigable in the first place. In some ways, the negative result in
Kleinberg's work is much stronger than the positive: for almost all
edge distributions efficient routing is not possible, it is only for
distributions meeting very strict criteria that it is. This seems
strange in relation to the lessons of Milgram's experiment -- people
really could route well -- and also Liben-Nowell's observations from
his dataset. It seems feasible that there is some dynamic which causes
navigability to arise. Sandberg and Clarke \cite{sandberg:evolving}
\cite{sandberg:neighbor} have suggested as such a dynamic a re-wiring
algorithm which causes networks to become navigable. By simulating a
large number of searches on the network, and changing the shortcuts
based on the path taken by each search, the algorithm progressively
creates a small-world from any starting distribution.

The goal of this paper is to make an experimental analysis of Sandberg
and Clarke's algorithm under more realistic situations. As far as we
are aware, this is the first comprehensive experimental analysis of a
dynamic model for the emergence of small worlds in realistic
geographic scenarios . We study how it behaves when vertices are not
placed in a grid, but rather distributed in a continuum and with
non-homogeneous population density. We contrast this with the results
of using the same edge probabilities as in the homogeneous case, as
well as the methods of Liben-Nowell et al.\ \cite{nowell:geographic}.
We also investigate how the algorithm responds to uneven distributions
in the source and destination of searches -- something more similar to
the power-law (``scale-free'') distributions known to be common in
many real life networks.

\subsection{Previous Work}

For a recent summary of previous work in the field of navigable
networks, see Jon Kleinberg's ICM survey \cite{kleinberg:networks}.
Besides the algorithm of Sandberg and Clarke studied here, Clauset and
Moore have suggested a different re-wiring algorithm which they find
experimentally also leads to a navigable graphs \cite{clauset:navig}.
The two methods are superficially similar, but actually lead to very
different dynamics. Other models for small-world emergence have been
suggested by Sandberg \cite{sandberg:double} and Duchon et al.\
\cite{duchon:emergence} and recently by Chaintreau et al.\
\cite{chaintreau:networks} but these are not evolutionary rewiring
schemes and function differently.

\subsection{Contribution}

We characterize our contribution as follows:

\begin{enumerate}
\item We investigate experimentally navigable small-world models with
  non-homogeneous population distributions, identifying when simply
  applying Kleinberg's method of adding shortcuts in the grid fails
  to produce navigability. The fact that this method fails in some
  cases has been observed before \cite{nowell:geographic} but when
  this happens is not fully understood. Here we show a family of
  distributions where the original model demonstrably fails to produce
  small worlds.

\item We simulate the placement of shortcuts in such environments
  using an evolutionary rewiring model. We are not aware of any
  previous studies of small-world emergence models in realistic
  geographic settings.  We demonstrate that the algorithm used produces
  navigable networks robustly under all tested circumstances: synthetic
  distributions with homogeneous and non-homogeneous distributions of data
  points, as well as scenarios based on the real world population
  distributions of Sweden and the United States.

\item We test the same evolutionary model also for non-homogeneous
  popularity models -- when some people are more popular targets for
  searches than others, for instance obeying various power laws.  We
  find that it not only works robustly well in these cases, but it
  ``learns'' the distribution and produces better mean results than
  otherwise. Such learning is not possible using the explicit
  augmentation models of Kleinberg \cite{kleinberg:smallworld} and
  Liben-Nowell \cite{nowell:geographic}.

\end{enumerate}

\section{Decentralized Routing and Navigable Augmentation}

Let $G = (V,E)$ belong to a family of finite graphs with high (some
power of $|V|$) diameter, and let the random graph $G'$ be created by
addition (augmentation) of random edges to $G$. It is well known, see
for instance \cite{bollobas:diameter}, that the diameter shrinks
quickly to a logarithm of $|V|$ when random edges are added between
the vertices.  Navigability concerns not a small diameter, however,
but rather a stronger property: the possibility of finding a short
path between two vertices in $G'$ using only local knowledge at each
vertex visited. By local knowledge, one means that each vertex knows
distance with respect to $G$, but does not know which random edges
have been added to any vertex until it is visited.  The exact limits
of such \emph{decentralized routing algorithms} have been discussed
elsewhere \cite{kleinberg:smallworld} \cite{barriere:efficient}, but
we will discuss only the one we use: \emph{greedy routing}.

In greedy routing for a target vertex $z$, the next vertex chosen is
always that neighbor which is closest to $z$ according to the distance
induced by $G$ (with some tie-breaking rule applied). Both the
original and augmented edges can be used, but because the choice is
only optimal with respect to $G$, the path discovered by greedy
routing will seldom be a minimal path in $G'$.

Kleinberg originally let $G$ be a $2$-dimensional $n \times n$-grid
and independently added shortcuts from each vertex to random
destinations. Each shortcut is added to $x \in V$ such that for $y \in
V$, and some $\alpha \geq 0$
\begin{equation}
  \PR(x \leadsto y) = {1 \over {h_{\alpha,n} d(x,y)^\alpha}}
  \label{eq:gridaug}  
\end{equation}
where $x \leadsto y$ is the event that $x$ is augmented with an edge
to $y$ and $d(x,y)$ denotes $L^1$ distance in $\Z^d$. $h_{\alpha,n}$
is here a normalizing constant, equal to $\sum_{y \neq x}
d(x,y)^{-\alpha}$. His observation was that when $\alpha = 2$, greedy
routing between any two points in $V$ takes $O(\log^2 n)$ steps in
expectation, while for any other value of $\alpha$ decentralized
algorithms create routes of expected length at least $\Omega(n^s)$
steps for some $s > 0$ (where $s$ depends on $\alpha$ and the
dimension but not size nor the algorithm used).

One may note that for $x$ in a 2-dimensional grid and $r > 0$, $|\{y
\in V : d(x, y) \leq r\}| \propto r^2$. The general principle that may
be noted by combining this with (\ref{eq:gridaug}) is that under
navigable augmentation the probability that $x$ links to $y$ should be
inversely proportional to the number of vertices that are closer to
$x$ than $y$.  This has been observed to hold not just when $G$ is a
grid of any dimension, but also for wider classes of graphs, see e.g.
\cite{kleinberg:dynamics} \cite{duchon:anygraph}.

In particular, in Liben-Nowell et al.'s paper on geographic routing
\cite{nowell:geographic}, they let the \emph{rank} of a vertex $y$
with respect to $x$ be $y$'s position when the vertices are ordered by
distance from $x$, written $\text{rank}_x(y)$. (Some natural ordering
of the vertices is used for tie-breaking). 
Their augmentation principle is then that
\begin{equation}
  \PR(x \leadsto y) = {1 \over h_n \text{rank}_x(y)}.
  \label{eq:rankaug}
\end{equation}
where $h_n = \sum_{k = 1}^n k^{-1} \approx \log n$. In a companion
paper by Kumar et al.\ \cite{kumar:geographic} they prove analytically
that this leads to a $O(\log^2 n)$ path lengths in expectation in a
discrete non-homogeneous model (the population is confined to a two
dimensional grid, but the number of people at each grid point varies).

\subsection{Continuum Settings}

When attempting to model reality, it is preferable to view the
``world'' of the vertices as a continuous metric space, rather than
just a base graph $G$. In particular, we want both the routing and the
augmentation to be with respect to the distance between vertices given
arbitrary positions in the space, rather than just graph distance.

That is, if $M$ is a metric space with $d : M \times M \mapsto \R$ a
metric, then the set $V$ may consist of $n$ randomly distributed
points in $M$. (Typically, and in the text below, the metric space is
a compact subset of $\R^2$, and $d$ is Euclidean distance.) We then
construct the ``short-range'' links (that is $G$) so as to respect the
geometry of the space. In particular, one wishes for $G$ to be
suitable for greedy routing with respect to $d$ in the sense that for
$z \neq x$, $x$ always has a neighbor closer to $z$ than itself -- if
this is not the case, it is possible for a greedy route to reach a
``dead-end'' at which no progress can be made in the next step.

This sort of construction was considered in \cite{draief:poisson}.
There, the authors let the the base graph $G$ be constructed by
connecting each vertex $x$ with all vertices within some distance
$r(n)$. For sufficiently large $r(n)$ this will with high probability
lead a base graph which is suitable for greedy routing. In
\cite{sandberg:neighbor} a different approach is used.  Instead of
connecting all near vertices, a Voronoi tessellation of $M$ with
respect to the points is calculated, and each vertex is connected to
those with neighboring cells. Thus $G$ is the Delaunay graph (or
Delaunay triangulation) of the set of points. The advantage of this
approach is that $G$ is a planar graph more elegantly describing a
neighbor structure on $M$, and that no probability calculations are
necessary: it is easy to see that $G$ always allows a greedy route to
monotonically approach its target. Delaunay graphs can be efficiently
calculated using well-known algorithms \cite{fortune:sweepline}.

  \begin{figure}
    \centering
   \includegraphics[width=7cm]{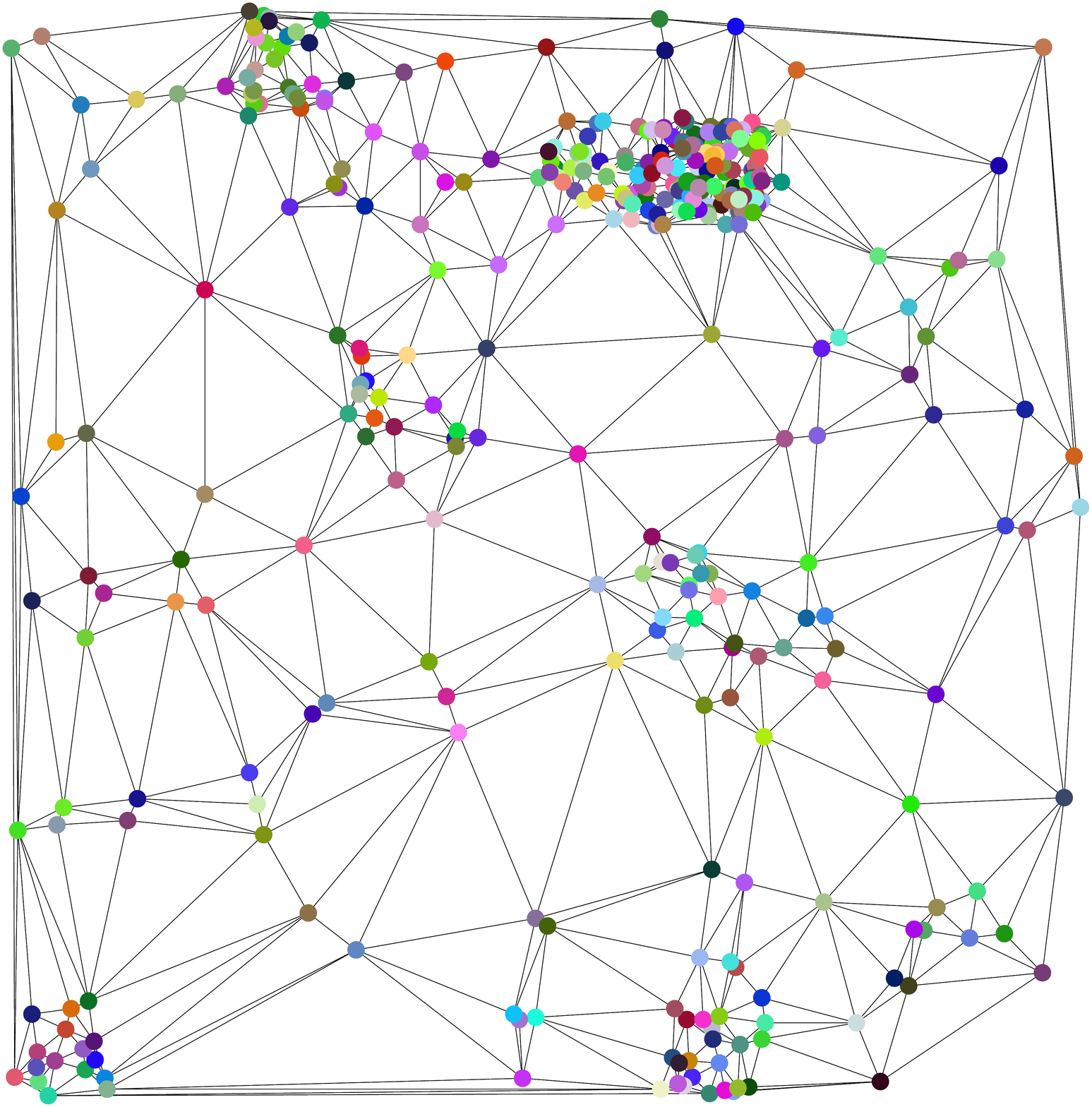}
   \caption{Realization of 375 vertices in a \emph{random}
     distribution (see Section \ref{sec:popdens}) on
     $[0,1]\times[0,1]$ using $k = 10$ and $\gamma = 1.2$, together
     with the Delaunay triangulation.}
    \label{fig:randomreal}
  \end{figure}

Once $G$ has been defined, one can augment it to create $G'$ as
before, adding outgoing edges to each vertex. The probabilities are
found by replacing $L^1$ distance with the more general $d(x, y)$ in
(\ref{eq:gridaug}) and when calculating $rank_{x}(y)$.

\subsection{Destination Sampling}

``Destination sampling'' is a name given to the re-wiring algorithm
introduced by Sandberg and Clarke in \cite{sandberg:evolving}. This is
not a method of augmenting a graph $G$ to create $G'$ as such, but
takes any given augmentation, and changes the shortcuts (without
changing their number or the out-degree of any vertex) so as to make
the resulting graph navigable.

The algorithm can be expressed in varying levels of generality, but
the general principle is always the same: each vertex samples the
destination of its shortcuts from among the destinations of searches
that pass through that vertex.

\begin{myalg} 
  Let $G_s = (V, E \cup E_s)$ be an augmented graph at time $s$. $G =
  (V, E)$ is the base graph, and $E_s$ the set of shortcuts, which for
  each vertex in $V$ contains at least one outgoing edge.

  Let $0 < p < 1$. Then $G_{s+1}$ is defined as follows.

\begin{enumerate}
\item Choose $y_{s+1}$ and $z_{s+1}$ randomly from $V$.
\item If the chosen vertices are distinct, do a greedy walk in $G_s$
  from $y_{s+1}$ to $z_{s+1}$. Let $x_0 = y_{s+1}, x_1, x_2, ..., x_t
  = z_{s+1}$ denote the points of this walk.
\item For each $x_0, x_1,...,x_{t-1}$ independently and with
  probability $p$ replace a randomly chosen shortcut from that vertex
  with one to $z_{s+1}$.
\end{enumerate}
\label{alg:ds}
\end{myalg}

  \begin{figure}[t]
    \centering
   \includegraphics[width=6cm]{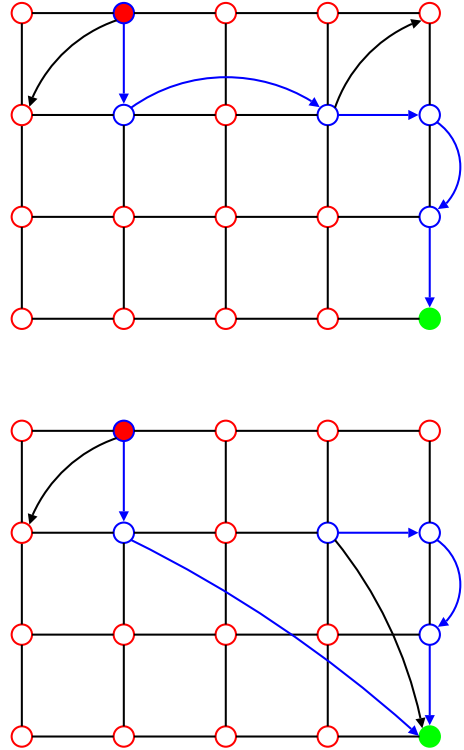}
   \caption{An illustration of Destination Sampling on an augmented
     grid before and after a rewiring. The blue vertices and edges
     represent a greedy path from red to green. After the path has
     been found, the shortcuts of two vertices along the path are
     randomly selected to be rewired.}
    \label{fig:destsamp}
  \end{figure}

In order to create a navigable augmentation, this algorithm is applied
repeatedly, causing it to converge to a limiting stationary
distribution. For simulations and analytical motivations why this
works when
\begin{enumerate}
\item The vertices of $V$ are homogeneously distributed.
\item $y_{s+1}$ and $z_{s+1}$ are chosen uniformly at random.
\end{enumerate}
see \cite{sandberg:evolving} and \cite{sandberg:neighbor}. The goal of
this paper is to see what happens when these two things do not
necessarily hold, as one would expect in a realistic situation.

The parameter $p$ in the algorithm is used to limit the dependence
among edges of nearby nodes. In theory, the algorithm performs better
the lower $p$ is, but the sampling, and thus convergence, is slower.
We use a value of $p = 0.1$ which we have determined experimentally
provides a good trade-off, throughout the paper.

\section{Experiments}

\subsection{Population Density}
\label{sec:popdens}
To experiment with non-homogeneous population densities we used a
continuum model and the Delaunay graph as described above. We
divided a 2-dimensional real space into zones of different population
density, and populated it with a non-homogeneous spatial Poisson
process. The intensity of the process in each zone was the zones
population density normalized so as to give approximately a desired
population size for the whole space.
With the vertices thus placed a Delaunay graph can then be constructed
using known algorithms, and we experimented with different ways of
augmenting edges to ensure navigability.

Our goals with this were twofold -- firstly to identify in which types
of situations augmenting according to formula (\ref{eq:gridaug}) fails
to lead to navigability, while (\ref{eq:rankaug}) does. Liben-Nowell
et al.\ \cite{nowell:geographic} give some hints as to when this may
be the case, but do not characterize it. Secondly, we tried applying
Destination Sampling (Algorithm \ref{alg:ds}) to see if it adapts and
gives navigable augmentation even in cases where straightforward
augmentation does not.

The models of population density that were used are as follows: 
\begin{enumerate}
\item \emph{Uniform}: The $n$ vertices are a placed uniformly at
  random across a square space $M = [0,1] \times [0,1]$.

\item \emph{Metropolis}: Here also the vertices are placed randomly in
  the same square space, but this time with 90\% of the total
  intensity within 20\% of the maximal distance from the space's
  center.



\item \emph{Random}: $[0,1] \times [0,1]$ was divided into $k \times
  k$ equally sized square zones, which were given a randomly ordered
  labeling of $s = 1,\hdots,k^2$. The population of each zone was then
  given a relative population density of $1/s^\gamma$, making the
  labels an ordering from most to least densely populated. For our
  experiments, we used $k = 100$ and $\gamma = 1.2$ where the latter
  value approximates the average value of decay of city sizes in the
  real world \cite{soo:zipflaw}. See Figure \ref{fig:randomreal} for
  an example realization.

\begin{figure}
  \begin{center}
    \includegraphics[width=3.8cm]{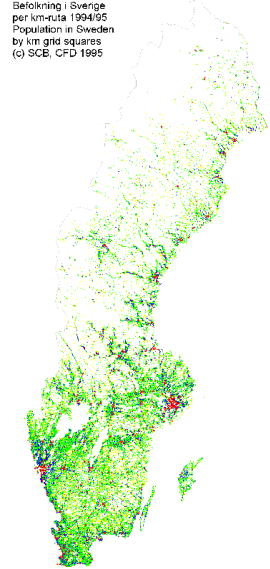} 
  \end{center}
  \caption{The population density of Sweden, broken into 1x1 km squares.}
  \label{fig:swedenpop}    
\end{figure}

\item \emph{Real World}: Finally, we used data regarding the
  contemporary population distribution of Sweden and the United
  States. For Sweden, data was obtained from Statistics Sweden
  \cite{scb:rutkarta} giving the population of each of the country's
  449,964 square kilometers, which we interpret as the proportional
  intensity of population in that area. For the United States, a map
  showing the population density of each of county in the lower 48
  states, taken from The National Atlas \cite{usgov:natatlas} was
  used\footnotemark{}.

  \footnotetext{The data in both cases was not exact. The map of
    Sweden gives the population of each square kilometer among the
    levels 0, 1-4, 5-29, 30-149, 150-4999, and 5000 and above. The map
    of the continental USA was divided in to 0, 1-4, 5-9, 10-24,
    25-49, 50-99, 100-249, and 250 and above people per square mile.
  }

\end{enumerate}

In each case, each vertex was given one outgoing shortcut, selected by
the following augmentations methods:

\begin{enumerate}
\item \emph{Distance}: Explicit sampling according to a power-law of
  the distance, as in Kleinberg's original work. This means following
  formula (\ref{eq:gridaug}) but with $d$ in the formula and the
  normalizer interpreted as Euclidean distance in $\R^2$. 

  For two dimensions, we use $\alpha = 2$, the value at which
  navigability arises in uniform networks.

\item \emph{Rank}: Explicit sampling, but using the rank formula
  (\ref{eq:rankaug}) as used by Liben-Nowell et al.

\item \emph{Destination Sampling}: Each node is initiated with no
  useful long-range link (formally, it has one to itself), and then
  the Algorithm \ref{alg:ds} is run $10 n$, where $n$ is the graph
  size, times.
\end{enumerate}
In some cases, we also compared with the results of choosing the
shortcut uniformly among the other vertices. This is known to give
greedy path lengths which are a fractional power of the number of
vertices -- $\Omega(n^{1/3})$ with a uniform population distribution
-- and thus was used only as a baseline for comparison.

\begin{figure}
    \begin{center}
    \includegraphics[width=3.8cm]{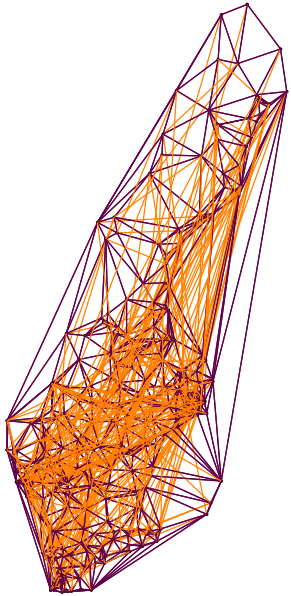} 
  \end{center}
  \caption{A realization of a Delaunay graph and destination sampling
    on a population distributed according to Sweden's population
    density (see Figure \ref{fig:swedenpop}).}
  \label{fig:swedgraph}
\end{figure}

\subsection{Popularity Distributions}

One of the most striking differences between social networks and most
simple random graph models is that the former seem to have power-law
degree distributions, while the latter most often have Poisson
distributed or even constant degrees. The celebrated ``preferential
attachment'' model (see \cite{barabasi:emergence} and
\cite{newman:structure} as well as \cite{bollobas:scalefree} for
rigorous analysis) explains this fact by showing that such
distributions arise when new vertices are more likely to connect to
vertices which already have a high degree.

In the context of the re-wiring algorithm, one might expect some
vertices to be more popular targets of searches than others (as would
most definitely be the case if the algorithm was, for example, used to
wire up a peer-to-peer network). It is of interest to see whether the
rewiring algorithm can adapt also to this situation. (Clearly the
distribution of searches has a large effect on what edges are formed
during the destination sampling.)

In order to separate these results from those above, we return to the
original model of Kleinberg -- vertices placed in a regular lattice with
connections to their nearest neighbors, as well as a single outgoing
long-range contact. The out-degree of each vertex is thus still fixed,
but the in-degree varies as a result of target popularity.

In our experiments, we produce a random order of the vertices, and
consider each vertex $x$'s position in this order, $p(x)$, as its
popularity ranking. We then select the targets of queries using
power-laws, of the form
\begin{equation}
\PR(\text{choosing }x\text{ as a target}) \propto p(x)^{-\beta}
  \label{eq:poppl}  
\end{equation}
for $\beta$ ranging from 0 to 2. As well as evaluating the performance
of the destination sampling under these conditions, we also study the
resulting degree distributions to see if a power-law is actually
recovered.


\subsection{Combined}

For completeness, we look at what happens when performing destination
sampling on the Swedish population model from the first section, while
at the same time using a biased popularity distribution as in the second.

\section{Results}

\subsection{Population Density}

\begin{figure}
    \begin{center}
    \includegraphics[width=7cm]{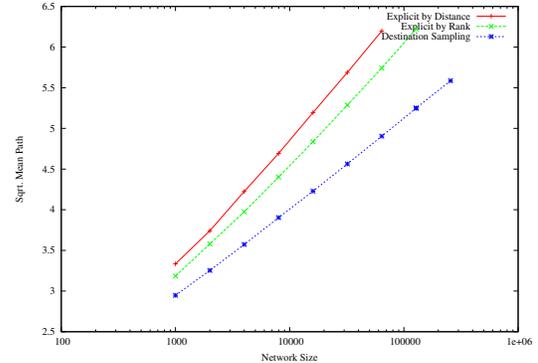} 
  \end{center}  
  \caption{Performance of greedy routing when augmenting the Delaunay
    graph of uniformly randomly distributed points in $[0,1] \times
    [0,1]$, using distance based augmentation (\ref{eq:gridaug}) with
    $\alpha = 2$, rank based
    augmentation (\ref{eq:rankaug}), and destination sampling.}
  \label{fig:unidens}
\end{figure}

\begin{figure}
    \begin{center}
    \includegraphics[width=7cm]{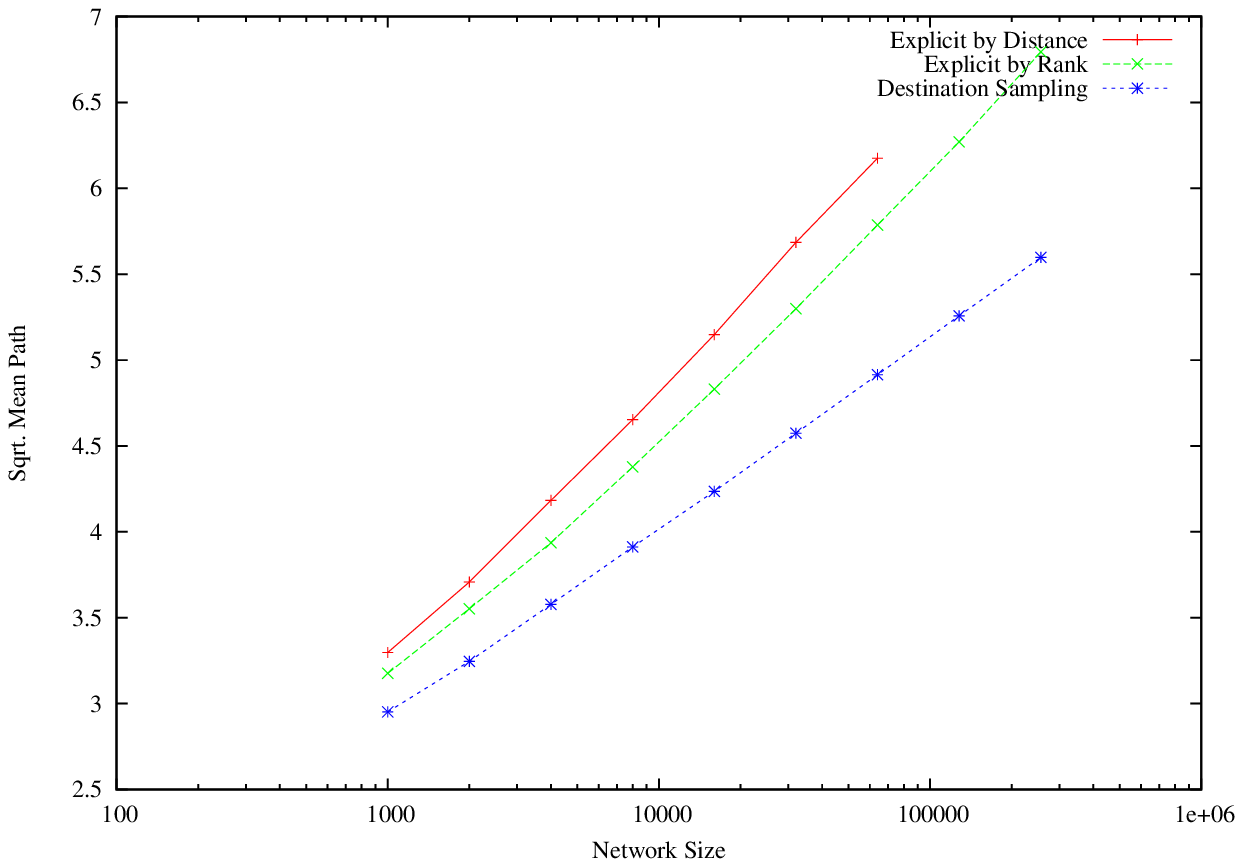} 
  \end{center}
  \caption{Performance of greedy routing when augmenting the Delaunay
    graph of \emph{Metropolis} distributed points in $[0,1] \times
    [0,1]$, using distance based augmentation augmentation
    (\ref{eq:gridaug}) with
    $\alpha = 2$, rank based augmentation (\ref{eq:rankaug}), and
    destination sampling.}
  \label{fig:metrodens}
\end{figure}

\begin{figure}
    \begin{center}
    \includegraphics[width=7cm]{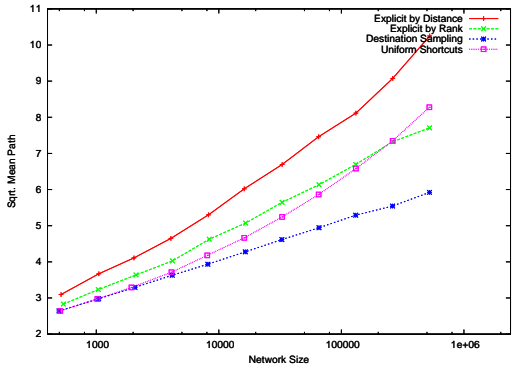} 
  \end{center}
  \caption{Performance of greedy routing when augmenting the Delaunay
    graph of \emph{Random} model distributed points in $[0,1] \times
    [0,1]$, using distance based augmentation augmentation
    (\ref{eq:gridaug}) with
    $\alpha = 2$, rank based augmentation (\ref{eq:rankaug}),
    destination sampling, and also choosing shortcuts uniformly.}
  \label{fig:randdens}
\end{figure}



\begin{figure}
    \begin{center}
    \includegraphics[width=7cm]{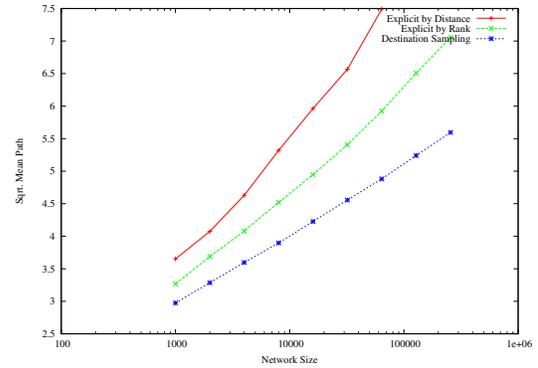} 
  \end{center}
  \caption{Performance of greedy routing when augmenting the Delaunay
    graph of points distributed according to Sweden's population,
    using distance based augmentation (\ref{eq:gridaug}) with $\alpha
    = 2$, rank based augmentation (\ref{eq:rankaug}), and destination
    sampling.}
  \label{fig:sweddens}
\end{figure}

\begin{figure}
    \begin{center}
    \includegraphics[width=7cm]{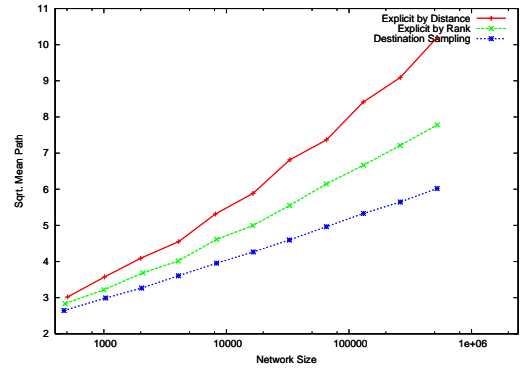} 
  \end{center}
  \caption{Performance of greedy routing when augmenting the Delaunay
    graph of points distributed according to the population density of
    the United States, using distance based augmentation
    (\ref{eq:gridaug}) with $\alpha = 2$, rank based augmentation
    (\ref{eq:rankaug}), and destination sampling.}
  \label{fig:usadens}
\end{figure}

\begin{figure}
    \begin{center}
    \includegraphics[width=7cm]{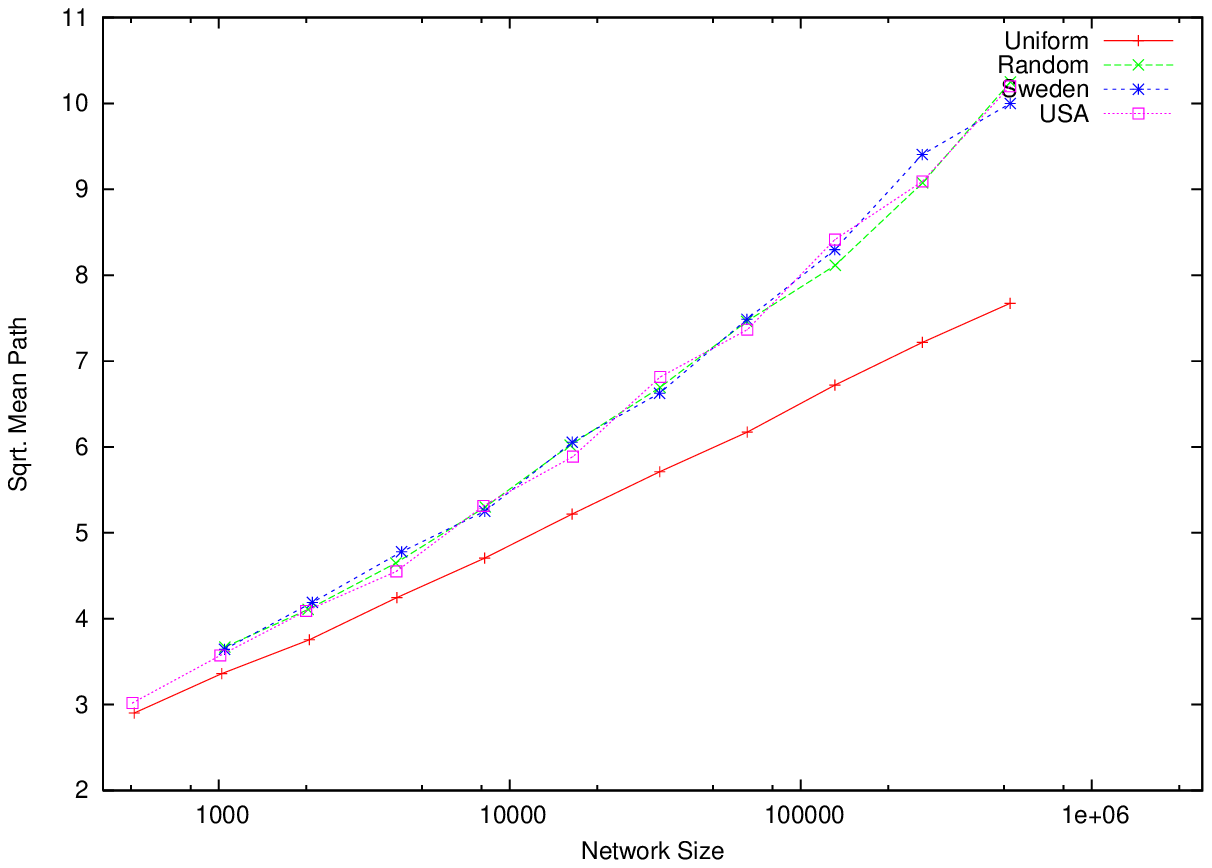} 
  \end{center}
  \caption{Explicit distance based augmentation gives the same (poor)
    performance in the Sweden, USA, and \emph{Random} population models.}
  \label{fig:bydist}
\end{figure}

Our results on non-homogeneous population distributions are shown in
Figures \ref{fig:unidens} -- \ref{fig:bydist}. In general, we find
that adding shortcuts as done by Kleinberg (\ref{eq:gridaug}) in the
grid can work well even when the population is not uniformly
distributed. This is shown by the fact that in the \emph{metropolis}
model, where most of the population is limited to a central core, we
still get $\log^2 n$ scaling of the path lengths, see Figure
\ref{fig:metrodens}. We find that in order for the purely distance
based augmentation to fail, we have to turn to highly irregular
population models. Previously it was known (see Liben-Nowell et al.\
\cite{nowell:geographic}) that the Kleinberg augmentation could fail,
but it was not clear under what circumstances. We identify one such
family, of irregular non--homogeneous distributions, namely the
\emph{random} model described above, where distance based augmentation
performs worse than even uniform such (Figure \ref{fig:randdens}).


Figures \ref{fig:sweddens} and \ref{fig:usadens} show the results
using the real world data of the population density in Sweden and the
USA. As expected due to Liben-Nowell et al.'s observations of the
Internet community data, as well as our structurally similar
\emph{random} configuration, purely distance based augmentation does
not give good result here, being beaten even by uniform augmentation.

Remarkably, the results for the \emph{random} model are almost
identical to those for the real world populations (Figure
\ref{fig:bydist}), showing that as far as distance based augmentation
is concerned, the real world data appears structurally very similar to
that produced by the random model.

The destination sampling algorithm performs well in \emph{all}
situations -- both synthetic test data as shown in Figures
\ref{fig:unidens} - \ref{fig:bydist} and for real world data, as
shown in Figures \ref{fig:sweddens} and \ref{fig:usadens} -- always
producing results that scale like $\log^2 n$ as desired, and
consistently performing better than explicit distance or rank based
augmentation. We have not been able to find \emph{any} population
distribution or situation where the re-wiring algorithm performs
poorly.

\subsection{Popularity Distributions}

\begin{figure}
    \begin{center}
    \includegraphics[width=7cm]{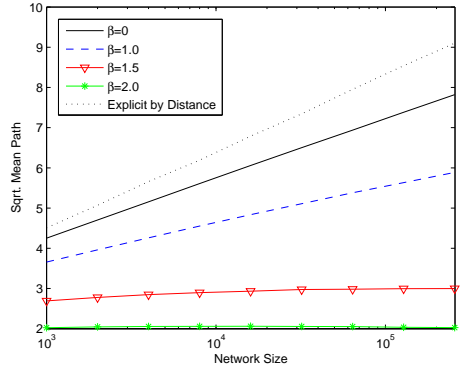} 
  \end{center}
  \caption{Performance of greedy routing when choosing destination by
    power-law distributions of different exponent ($\beta$ in
    (\ref{eq:poppl})), using destination sampling in a one dimensional
    grid. This is contrasted against explicit augmentation by distance
    (with $\alpha = 1$), where the destination distribution of course makes
    no difference.}
  \label{fig:popdists}
\end{figure}

\begin{figure}
  \begin{center}
    \includegraphics[width=7cm]{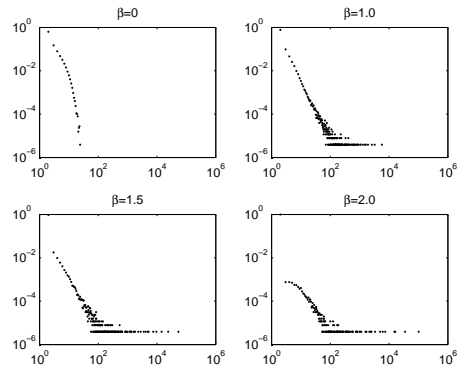}
  \end{center}
  \caption{Degree distributions of the graphs created by destination
    sampling when choosing choosing destinations by power-law, for
    different values of the exponent ($\beta$) in (\ref{eq:poppl}).
    the network has 256,000 vertices, each with out-degree one. The
    plots show the fraction of vertices with each total in-degree
    (rounded up to the nearest multiple of ten).}
  \label{fig:popdegdist}
\end{figure}

Figures \ref{fig:popdists} -- \ref{fig:popdegdist} show our results
when we return to a homogeneous grid, but instead let the popularity
of vertices as destinations vary. As in the cases above, we find the
destination sampling excels, producing shorter paths than explicit
augmentation. In fact, as $\beta$ increases in (\ref{eq:poppl}), we
find that destination sampling gives shorter and shorter paths.
Intuitively, this follows from the fact that since most searches are
going to a limited set of vertices, most shortcuts also lead to those
vertices, allowing most of the routing to occur within a small subset
of the whole graph. This is most clear when $\beta > 1$, in which case
one can for any $\epsilon > 0$ fix an $m$ such that at least $1 -
\epsilon$ of the queries are destined for the $m$ most popular
vertices, independent of the graph size $n$. Indeed, one can see in
Figure \ref{fig:popdists} that we observe no increase in the path
lengths for $\beta = 2$ after a certain point. In contrast, both the
distance based augmentation as well as Liben-Nowell et al.\ rank based
augmentation, which assign fixed probabilities independently of the
popularity distribution, of course do not take advantage of the
non-uniform popularity distribution at all.

The resulting degree distributions are described in Figure
\ref{fig:popdegdist}. One can see that when the destination is
selected according to a power-law distribution, the degree of the
vertices also end up following such a distribution. This is not
particularly surprising, given the way the algorithm functions, but
shows that we can generate power-law (``scale-free'') graphs without
sacrificing navigability.

\subsection{Combined}

\begin{figure}
    \begin{center}
    \includegraphics[width=7cm]{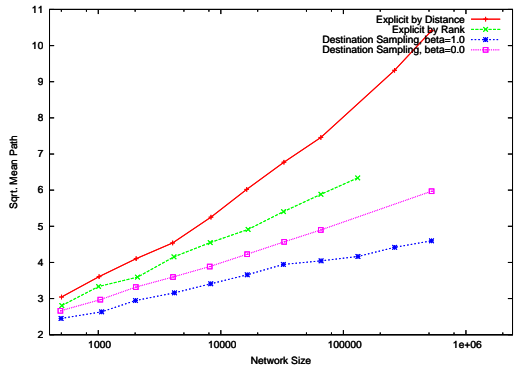} 
  \end{center}
  \caption{Performance of greedy routing when combining the
    \emph{Sweden} population distribution with a power-law popularity
    of the destinations ($\beta$ as in (\ref{eq:poppl})).}
\label{fig:combined}
\end{figure}

As expected from the above results, destination sampling functions
well also when combining both a non-homogeneous geographical
population density, and a power-law distribution of destination, see
Figure \ref{fig:combined}. The mean path length is considerably
shorter than that attained under either the distance or rank based
augmentations.

\section{Conclusion}

We find, as has been observed before, that in geographic networks the
augmentation process is sensitive to the environment and distribution
of the vertices. If the distribution deviates sufficiently in
structure from a uniform placement of the vertices, otherwise
effective methods of assigning shortcuts will not work. The
destination sampling algorithm, however, is adaptive and will create
navigable graphs in all cases that we have studied.

Likewise, when the popularity of the different vertices as
destinations for searches is uneven, the destination sampling
algorithm will adapt and is able to utilize this to find even shorter
paths.

Given its remarkable ability to create augmentations reflective of
each situation, we believe that in any case where navigable
augmentation is possible (see \cite{duchon:anygraph}
\cite{fraigniaud:doubling} for a discussion on the limits of this)
destination sampling can be used to achieve one. We note that the
formulation of the algorithm requires no understanding of the actual
situation - the same exact procedure will work regardless of
geographic or popularity distribution, which is not true, for
instance, when augmenting by rank. 




\bibliographystyle{hunsrt}
\bibliography{../../tex/ossa}

\end{document}